\documentclass[a4paper,12pt]{article}
\usepackage{epsfig}
\usepackage[dvips,usenames]{color}
\usepackage{graphicx}

\newlength{\dinwidth}
\newlength{\dinmargin}
\begin{document}
\title{The correction of the littlest Higgs model to the Higgs
production process $e^{-}\gamma\rightarrow \nu_{e}W^{-}H$ in
$e^{-}\gamma$ collisions}
\bigskip
\author{Yao-Bei Liu$^{a}$, Lin-lin Du$^{b}$, Xue-Lei Wang$^{b}$ \\
{\small a: Henan Institute of Science and Technology, Xinxiang
453003, P.R.China}
\thanks{E-mail:hnxxlyb2000@sina.com}\\
 {\small b: College of Physics and Information
Engineering,}\\
\small{Henan Normal University, Xinxiang  453007, P.R.China}\\
 }
\maketitle
\begin{abstract}
\indent The littlest Higgs model is the most economical one among
various little Higgs models. In the context of the littlest
Higgs(LH) model, we study the process $e^{-}\gamma\rightarrow
\nu_{e}W^{-}H$ and calculate the contributions of the LH model to
the cross section of this process. The results show that, in most of
parameter spaces preferred by the electroweak precision data, the
value of the relative correction is larger than $10\%$. Such
correction to the process $e^{-}\gamma\rightarrow \nu_{e}W^{-}H$ is
large enough to be detected via $e^{-}\gamma$ collisions in the
future high energy linear $e^{+}e^{-}$ collider($LC$) experiment
with the c.m energy $\sqrt{s}$=500 GeV and a yearly integrated
luminosity $\pounds=100fb^{-1}$, which will give an ideal way to
test the model.
\end{abstract}
PACS number(s): 12.60Cn,14.70.Pw,14.80.Bn,13.66.Hk
\newpage
\section{Introduction}
\indent The standard model($SM$) provides an excellent effective
field theory description of almost all particle physics experiments.
But in the $SM$ the Higgs boson mass suffers from an instability
under radiative corrections. The naturalness argument suggests that
the cutoff scale of the SM is not much above the electroweak scale:
New physics will appear around TeV energies. The possible new
physics scenarios at the TeV scale might be
supersymmetry\cite{supersymmetry}, dynamical symmetry
breaking\cite{dynamical}, extra dimensions\cite{extra}. Recently, a
new model, known as little Higgs model has drawn a lot of interest
and it offers a very promising solution to the hierarchy problem in
which the Higgs boson is naturally light as a result of nonlinearly
realized symmetry \cite{little-1,little-2,little-3,littlest}. The
key feature of this model is that the Higgs boson is a
pseudo-Goldstone boson of an approximate global symmetry which is
spontaneously broken by a VEV at a scale of a few TeV and thus is
naturally light. The most economical little Higgs model is the
so-called
 littlest Higgs model, which is based on a $SU(5)/SO(5)$
 nonlinear sigma model \cite{littlest}. It consists of a $SU(5)$ global
 symmetry, which is spontaneously broken down to $SO(5)$ by a vacuum
 condensate $f$. In this model, a set of new heavy gauge bosons$(B_{H},Z_{H},W_{H})$ and
 a new heavy-vector-like quark(T) are introduced which just cancel
 the quadratic divergence induced by the $SM$ gauge boson loops and the
 top quark loop, respectively. The distinguishing features of this
 model are the existence of these new particles and their
 couplings to the light Higgs. The measurement of these new particle effects might
 prove the existence of the littlest Higgs mechanism.\\
 \indent  The hunt for the Higgs boson and the elucidation of the mechanism of symmetry
 breaking is one of the most important goals for present and future
 high energy collider experiments. Precision electroweak measurement
  data and direct searches suggest that the Higgs boson must be relative light and its mass should
 be roughly in the range of 114.4 GeV$\sim$208 GeV at $95\%$ CL \cite{Higgs}.
 The high energy linear $e^{+}e^{-}$ colliders($LC$) has a large potential for the discovery of new particles\cite{LC}.
 Due to its rather clean environment, the $LC$ will be perfectly
 suited for precise analysis of physics beyond the $SM$ as well as for
 testing the $SM$ with an unprecedented accuracy. An unique feature
 of the $LC$ is that it can be transformed to $\gamma\gamma$ or
 $e\gamma$ colliders with the photon beams generated by laser-scattering
 method.
 Their effective luminosity and energy are expected to be comparable
 to those of the $LC$. In some scenarios, they are the best
 instrument for the discovery of signatures of new physics. \\
 \indent Some
 phenomenological studies of the littlest Higgs model via $e\gamma$
 or $\gamma\gamma$ collision has been done \cite{ey}.
 The main W boson production mechanism is provided by the process
$\gamma$+e$\rightarrow$W+$\nu$. The larger cross section for W boson
production suggests that the process $e^{-}\gamma\rightarrow
\nu_{e}W^{-}H$ can be exploited as a source for Higgs boson emitted
from the W-boson line. In the context of the SM, this process has
been studied at leading order \cite{eysm}. Since the final state
consists of three particles two of which are heavy, the cross
section at moderate energies($\sqrt{s}\sim500GeV$) is smaller than
that of the standard $WW$ fusion mechanism $e^{+}e^{-}\rightarrow
\nu\bar{\nu}H$ , the Higgs-strahlung process
 $e^{+}e^{-}\rightarrow ZH$ and the $ZZ$ fusion mechanism $e^{+}e^{-}\rightarrow e^{+}e^{-}H$.
  These three
processes have been studied in the context of the SM\cite{ee-SM,eeH}
and the littlest Higgs model\cite{ee-LH}. However, at higher
energies the cross
 section for the process $e^{-}\gamma\rightarrow
\nu_{e}W^{-}H$ is nearly as large as that of the dominant
$\nu\bar{\nu}H$ and the process for most of the mass of Higss boson
range accessible at linear colliders, and significantly larger than
the cross section of the Higgs-strahlung process. A dedicated
$e^{-}\gamma$ collider with back scattered laser beam therefore
gives rise to a large Higgs production cross section through the
process $e^{-}\gamma\rightarrow \nu_{e}W^{-}H$. Thus, it is very
interesting to study this process in the popular specific models
beyond the SM. The purpose of this paper is to calculate the
corrections of new particles predicted by the littlest Higgs model
to the process $e^{-}\gamma\rightarrow \nu_{e}W^{-}H$ and see
whether the effects on this process can be observed in the future
$LC$ experiments with the c.m energy $\sqrt{s}$=500 GeV.

\indent  This paper is organized as follows. In section two, the
littlest model is briefly introduced, and then the production
amplitude of the process is given. The numerical results and
discussions are presented in section three. The conclusions are
given in section four.

\section{The littlest Higgs model and the production amplitude of $e^{-}\gamma\rightarrow \nu_{e}W^{-}H$
}
 \indent The littlest Higgs model is based on the
$SU(5)/SO(5)$ nonlinear sigma model. At the scale $\Lambda_{s}\sim
4\pi$$f$, the global $SU(5)$ symmetry is broken into its subgroup
$SO(5)$ via a vacuum condensate $f$, resulting in 14 Goldstone
bosons. The effective field theory of these Goldstone bosons is
parameterized by a non-linear $\sigma$ model with gauged symmetry
$[SU(2)\times U(1)]^{2}$, spontaneously broken down to its diagonal
subgroup $SU(2)\times U(1)$, identified as the SM electroweak gauge
group. Four of these Goldstone bosons are eaten by the broken gauge
generators, leaving 10 states that transform under the SM gauge
group as a doublet H and a triplet $\Phi$. This breaking scenario
also gives rise to four massive gauge bosons $B_{H}$,$Z_{H}$ and
$W^{\pm}_{H}$, which might produce the characteristic signatures in
the present and future high energy
collider experiments \cite{signatures-1,signatures-2,signatures-3}.\\
 \indent  After the electroweak symmetry breaking, the mass eigenstates
 are obtained via mixing between the heavy and light gauge bosons.
 They include the light (SM-like) bosons $Z_{L}$, $A_{L}$ and
$W^{\pm}_{L}$ observed at experiments, and new heavy bosons $Z_{H}$,
$B_{H}$ and $W^{\pm}_{H}$ that could be observed in the future
experiments. To obtain our numerical results, we write the masses of
the relevant particles as\cite{signatures-1}:
\begin{eqnarray}
M^{2}_{W_{L}}&=&(m_{W})^{2}\{1-\frac{v^{2}}{f^{2}}[\frac{1}{6}+\frac{1}{4}(c^{2}-s^{2})^{2}+
\frac{x^2}{4}]\},\\
M^{2}_{W_{H}}&\approx&(m_{W})^{2}(\frac{f^{2}}{s^{2}c^{2}v^{2}}-1),
\end{eqnarray}
with $x=\frac{4fv'}{v^{2}}$, where $m_{W}=ev/2s_{W}$ is the mass of
the SM gauge boson W, $v$=246 GeV is the elecroweak scale, $v'$ is
the vacuum expectation value of the scalar $SU(2)_{L}$ triplet and
$s_{W}(c_{W})$ represents the sine(cosine) of the weak mixing angle.
We define $x=4fv'/v^{2}$ to parametrize this vacuum expectation
value of the scalar triplet field $\phi$. The mass of neutral scalar
boson $M_{\phi^{0}}$ can be given as \cite{signatures-1}
\begin{eqnarray}
M^{2}_{\phi^{0}}=\frac{2m^{2}_{H^{0}}f^{2}}{v^{2}[1-(4v'f/v^{2})^{2}]}=\frac{2m^{2}_{H^{0}}f^{2}}{v^{2}(1-x^{2})}
\end{eqnarray}
The above equation about the mass of $\Phi$ requires a constraint of
0$\leq$x$<$1 (i.e.,$4v'f/v^{2}<1$), which shows the relation
between the scale $f$ and the vacuum expectation values of the Higgs field doublet and the triplet$(v,v')$.\\
\indent  Taking account of the gauge invariance of the Yukawa
coupling and the $U(1)$ anomaly cancelation, the couplings of the
relevant couplings of the gauge boson $W^{\pm}_{L}$ and
$W^{\pm}_{H}$ to ordinary particles and the Higgs boson can be
written as follows in the LH model \cite{signatures-1}:
\begin{eqnarray}
g_{V}^{W_{L}e\nu}&=&-g_{A}^{We\nu}=\frac{ie}{2\sqrt{2}s_{W}}[1-\frac{v^{2}}
{2f^{2}}c^{2}(c^{2}-s^{2})],\\
g_{V}^{W_{H}e\nu}&=&-g_{A}^{W_{H}e\nu}=-\frac{ie}{2\sqrt{2}s_{W}}\frac{c}{s},\\
g^{W^{+}_{L\mu}W^{-}_{L\nu}H}&=&\frac{ie^{2}}{2s^{2}_{W}}g_{\mu\nu}(1-\frac{v^{2}}{3f^{2}}+
\frac{1}{2}(c^{2}-s^{2})^{2}\frac{v^{2}}{f^{2}}-\frac{3vx}{f}),\\
g^{W^{+}_{L\mu}W^{-}_{H\nu}H}&=&\frac{-ie^{2}}{2s^{2}_{W}}\frac{(c^{2}-s^{2})}{2sc}vg_{\mu\nu}.
\end{eqnarray}
we write the gauge boson-fermion couplings in the form of
 $i\gamma^{\mu}(g_{V}+g_{A}\gamma^{5})$.
With all momenta out-going, the three-point gauge boson
self-couplings can be written in the form of:
\begin{eqnarray}
V_1^{\mu}(k_1)V_2^{\nu}(k_2)V_3^{\rho}(k_3):~~
-ig_{V_1V_2V_3}[g^{\mu\nu}(k_1-k_2)^{\rho}+g^{\nu\rho}(k_2-k_3)^{\mu}+g^{\rho\mu}(k_3-k_1)^{\nu}],
\end{eqnarray}
 The coefficients $g_{V_1V_2V_3}$ are given as:
\begin{eqnarray}
g_{A_LW^+_LW^-_L}=g_{A_LW^+_HW^-_H}=-e,\hspace{2.7cm}
g_{A_LW^+_LW^-_H}=0,
\end{eqnarray}
 \indent  Compared with the process $e^{-}\gamma\rightarrow \nu_{e}W^{-}H$ in the SM, this process in the LH model
 receives additional contributions from the heavy boson $W^{\pm}_{H}$, proceed through the Feynman diagrams
 depicted in Fig1. Furthermore, the modification of the relations
 among the SM parameters, the precision electroweak input
 parameters, the correction terms to the SM We$\nu_{e}$ and
 $WWH$ coupling can also produce corrections to this
 process.\\
\begin{figure}[t]
\begin{center}
\epsfig{file=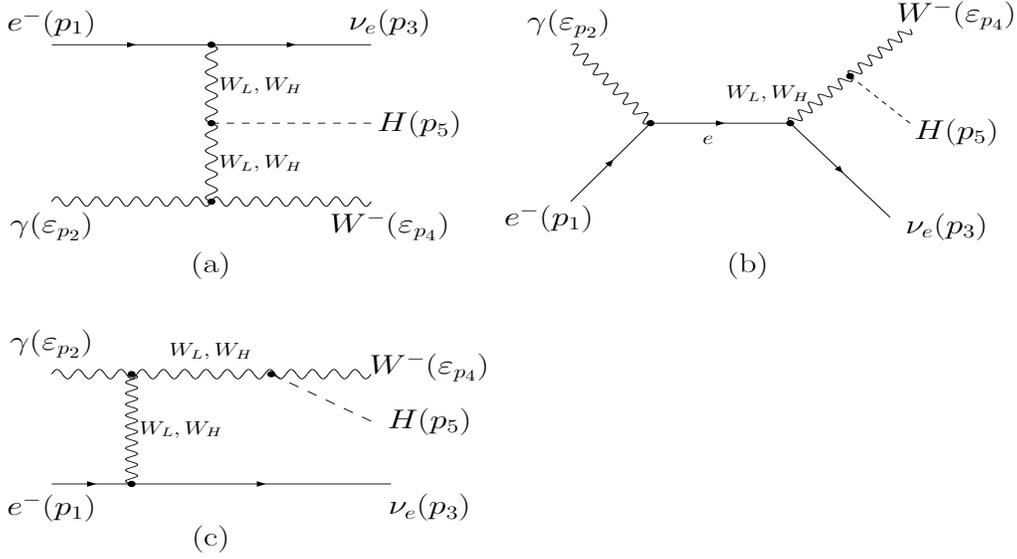,width=450pt,height=500pt} \vspace{-9cm}
\caption{\small Feynman diagrams of the process
$e^{-}\gamma\rightarrow \nu_{e}W^{-}H$ in the littlest Higgs model.}
\label{fig1}
\end{center}
\end{figure}
 \indent In order to write a compact expression for the amplitudes, it is
 necessary to define the triple-boson couplings coefficient as:
\begin{equation}
 \Gamma^{\alpha\beta\gamma}(p_{1},p_{2},p_{3})=g^{\alpha\beta}(p_{1}-p_{2})^{\gamma}
 +g^{\beta\gamma}(p_{2}-p_{3})^{\alpha}+g^{\gamma\alpha}(p_{3}-p_{1})^{\beta},
 \end{equation}
 with all motenta out-going. The invariant production amplitudes of the process
 can be written as:
\begin{equation}
 M=M^{a}+M^{b}+M^{c},
 \end{equation}
 with
 \begin{eqnarray*}
 M_{a}&=&[AG^{\mu\nu}(p_{1}-p_{3},M_{W_{L}})+BG^{\mu\nu}(p_{1}-p_{3},M_{W_{H}})]\overline{u}_{\nu}(p_{3})\gamma^{\mu}(1-\gamma^{5})u_{e}(p_{1})\\
& &G^{\nu\alpha}(p_{2}-p_{4},M_{W_{L}})\Gamma^{\alpha\rho\sigma}(p_{2}-p_{4},-p_{2},p_{4})\varepsilon^{\rho}(p_{2})\varepsilon^{\sigma}(p_{4})\\
 M_{b}&=&[AG^{\mu\nu}(p_{4}+p_{5},M_{W_{L}})+BG^{\mu\nu}(p_{4}+p_{5},M_{W_{H}})]G(p_{1}+p_{2})\\
 &
 &\overline{u}_{\nu}(p_{3})\gamma^{\mu}(1-\gamma^{5})\gamma^{\rho}u_{e}(p_{1})\varepsilon^{\rho}(p_{2})\varepsilon^{\nu}(p_{4})\\
 M_{c}&=&[AG^{\mu\nu}(p_{1}-p_{3},M_{W_{L}})G^{\rho\sigma}(p_{4}+p_{5},M_{W_{L}})+BG^{\mu\nu}(p_{1}-p_{3},M_{W_{H}})G^{\rho\sigma}(p_{4}+p_{5},M_{W_{H}})]\\
 & &\overline{u}_{\nu}(p_{3})\Gamma^{\nu\alpha\rho}(p_{2}-p_{4}-p_{5},-p_{2},p_{4}+p_{5})\gamma^{\mu}(1-\gamma^{5})u_{e}(p_{1})\varepsilon^{\alpha}(p_{2})\varepsilon^{\sigma}(p_{4})
 \end{eqnarray*}
 with
\begin{eqnarray}
 A&=&\frac{e^{4}v}{4\sqrt{2}s^{3}_{W}}[1-\frac{v^{2}}
{2f^{2}}c^{2}(c^{2}-s^{2})]\{1-\frac{v^{2}}{f^{2}}[\frac{1}{3}-
\frac{1}{2}(c^{2}-s^{2})^{2}+\frac{3fx}{v}]\},\\
 B&=&\frac{e^{4}v}{8\sqrt{2}s^{3}_{W}}\frac{(c^{2}-s^{2})}{s^{2}}
 \end{eqnarray}
 Here, $G^{\mu\nu}(p,M)=\frac{-ig^{\mu\nu}}{p^{2}-M^{2}}$ is the propagator of the
 particle.

  We can see that one source of the corrections of the
 littlest Higgs model to the process arises from the new heavy gauge bosons $W^{\pm}_{H}$. On the other hand, the littlest Higgs model can generate the correction to the mass
 of gauge boson W in the SM and to the tree-level coupling vertices, which can also produce the correction to the process.
 In our numerical calculation, we will also take account of such correction effect. \\
\indent The hard photon beam of the $e\gamma$ collider can be
obtained from laser backscattering at the $e^{+}e^{-}$ linear
collider. Let $\hat{s}$ and $s$ be the center-of-mass energies of
the $e\gamma$ and $e^{+}e^{-}$ systems, respectively. After
calculating the cross section $\sigma(\hat{s})$ for the subprocess
$e^{-}\gamma\rightarrow \nu_{e}W^{-}H$, the total cross section at
the $e^{+}e^{-}$ linear collider can be obtained by folding
$\sigma(\hat{s})$ with the photon distribution function that is
given in Ref \cite{function}:
\begin{equation}
\sigma(tot)=\int^{x_{max}}_{(M_{W}+M_{H})^{2}/s}dx\sigma(\hat{s})f_{\gamma}(x),
 \end{equation}
where
\begin{equation}
f_{\gamma}(x)=\frac{1}{D(\xi)}[1-x+\frac{1}{1-x}-\frac{4x}{\xi(1-x)}+\frac{4x^{2}}{\xi^{2}(1-x)^{2}}],
\end{equation}
with
\begin{equation}
D(\xi)=(1-\frac{4}{\xi}-\frac{8}{\xi^{2}})\ln(1+\xi)+\frac{1}{2}+\frac{8}{\xi}-\frac{1}{2(1+\xi)^{2}},
\end{equation}
In the above equation, $\xi=4E_{e}\omega_{0}/m_{e}^{2}$ in which
$m_{e}$ and $E_{e}$ stand, respectively, for the incident electron
mass and energy, $\omega_{0}$ stands for the laser photon energy,
and $x=\omega/E_{e}$ stands for the fraction of energy of the
incident electron carried by the backscattered photon. $f_{\gamma}$
vanishes for $x>x_{max}=\omega_{max}/E_{e}=\xi/(1+\xi)$. In order to
avoid the creation of $e^{+}e^{-}$ pairs by the interaction of the
incident and backscattered photons, we require
$\omega_{0}x_{max}\leq m_{e}^{2}/E_{e}$, which implies that $\xi\leq
2+2\sqrt{2}\simeq4.8$. For the choice of $\xi=4.8$, we obtain
\begin{equation}
x_{max}\approx0.83,~~~~~~~~~~~~~~~~~~~D(\xi_{max})\approx1.8.
 \end{equation}
 For simplicity, we have ignored the possible polarization for the
 electron and photon beams.\\
\indent In the calculation of $\sigma(\hat{s})$, instead of
calculating the square of the amplitudes analytically, we calculate
the amplitudes numerically by using the method of the
references\cite{hz}. This greatly simplifies our calculation.

\section{ The numerical results and discussions}
\indent In the LH model, the relation among the Fermi coupling
constant $G_{F}$, the
 gauge boson W mass $M_{W}$ and the fine structure constant $\alpha$
 can be written as\cite{gf}:
\begin{eqnarray}
\frac{G_{F}}{\sqrt{2}}=\frac{\pi\alpha}{2M^{2}_{W}s^{2}_{W}}[1-c^{2}(c^{2}-s^{2})\frac{v^{2}}{f^{2}}
+2c^{4}\frac{v^{2}}{f^{2}}-\frac{5}{4}(c'^{2}-s'^{2})\frac{v^{2}}{f^{2}}]
 \end{eqnarray}
 So we have
\begin{eqnarray}
\frac{e^{2}}{s^{2}_{W}}=\frac{4\sqrt{2}G_{F}M^{2}_{W}}{[1-c^{2}(c^{2}-s^{2})\frac{v^{2}}{f^{2}}+2c^{4}\frac{v^{2}}{f^{2}}-\frac{5}{4}(c'^{2}-s'^{2})\frac{v^{2}}{f^{2}}]}
 \end{eqnarray}

In the following numerical calculation, we take the input parameters
as
 $G_{F}=1.16637\times10^{-5}GeV^{-2}$, $M_{Z}^{SM}$=91.18 GeV, $s_{W}^{2}$=0.2315 and $M_{W}$=80.45GeV\cite{data}.  For the light Higgs boson H, in this paper,
 we only take the illustrative value $M_{H}$=120GeV. The value of the relative correction
parameter is insensitive to the degree of the electron and positron
polarization and the c.m. energy $\sqrt{s}$. Therefore, we do not
consider the polarization of the initial states and take
$\sqrt{s}$=500 GeV in our numerical calculation. There are four
parameters, $f$, c, $c'$, x, involved in the expression of the
relative correction parameter $\delta\sigma/\sigma^{SM}$ with
$\delta\sigma=| \sigma^{tot}-\sigma^{SM}|$ and $\sigma^{SM}$ is the
tree-level cross section of  $e^{-}\gamma\rightarrow \nu_{e}W^{-}H$
 production predicted by the SM. In the LH model, the custodial
$SU(2)$ global symmetry is
 explicitly broken, which can generate large contributions to the
 electroweak observables. However, if we carefully adjust the $U(1)$ section of the theory
 the contributions to the electroweak observables can be reduced and
 the constraints become
 relaxed. The scale parameter $f=1\sim2$ TeV is allowed for the mixing
 parameters $c$ and $c^{'}$ in the ranges of
 $0 \sim 0.5,0.62 \sim 0.73$ \cite{constraints}.
 In order to obtain the correct $EWSB$ vacuum and avoid
giving a TeV-scale $VEV$
 to the scalar triplet $\phi$, we should have that the value of
 parameter
$x=4fv'/v^{2}$ is smaller than 1\cite{signatures-1,v1}. The
numerical results are summarized in Figs.(2-4)

\begin{figure}[h]
\begin{center}
\scalebox{0.85}{\epsfig{file=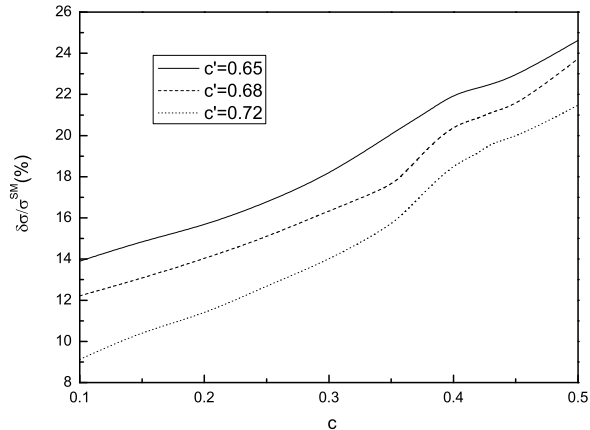}}\\
\end{center}
\caption{\small The relative correction $\delta\sigma/\sigma^{SM}$
as a function of the mixing parameter c for f=1 TeV, x=0.1 and three
values of the mixing parameter $c'$.}
\end{figure}

\indent  The relative correction $\delta\sigma/\sigma^{SM}$ is
plotted in Fig.2 as a function of the mixing parameter c for $f$=1
TeV, $x$=0.1 and $c^{'}=0.65, 0.68, 0.72$ respectively. From Fig.2,
we can see that the relative correction $\delta\sigma/\sigma^{SM}$
increases with the mixing parameter c increasing and sensitive to
the mixing parameter $c'$. For x($4fv'/v^{2}$)=0.1, the
 value of the relative correction $\delta\sigma/\sigma^{SM}$
is larger than $9\%$ in all of the parameter space preferred by the
electroweak precision data. When the mixing parameter c gets close
to 0.5, the value of the relative correction
$\delta\sigma/\sigma^{SM}$ is larger than $20\%$ in most of the
parameter space in the LH model.
\\
\begin{figure}[b]
\begin{center}
\scalebox{0.85}{\epsfig{file=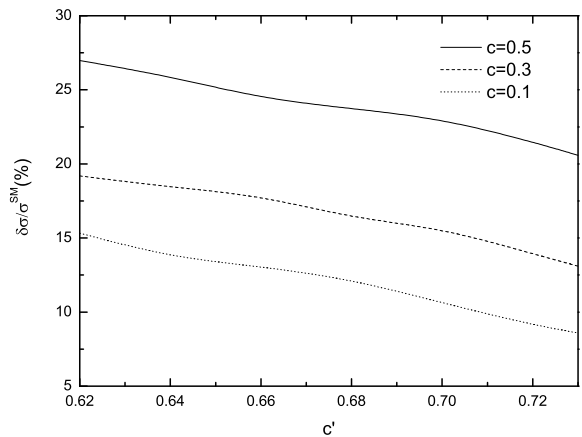}}
\end{center}
\caption{\small The relative correction $\delta\sigma/\sigma^{SM}$
as a function of the mixing parameter $c^{'}$ for f=1 TeV, $x=0.1$
GeV and c=0.1(dotted line), 0.3(dashed line) and 0.5(solid line).}
\end{figure}
 \indent To see the dependence of relative correction on the parameter $c'$, in Fig.3,
we plot $\delta\sigma/\sigma^{SM}$ as a function of the mixing
parameter $c'$ for $f$=1TeV, $x$=0.1, and three values of the mixing
parameter c. We can see that the relative correction decreases
slowly as the mixing parameter $c'$ and is also sensitive to the
mixing parameter c. In most of the parameter space of the LH model,
the value of $\delta\sigma/\sigma^{SM}$ is larger than $10\%$, which
might be detected in the future LC experiments.

\indent In Fig.4, we plot $\delta\sigma/\sigma^{SM}$ as a function
of the parameter $x$($x=4fv'/v^{2}<1$) for three values of the scale
parameter $f$($f$=1, 1.5, 2TeV) and take $c=0.3$, $c^{'}=0.68$. One
can see that the relative correction increases with the value of
parameter $x$ increasing. This is because the contribution of
littlest Higgs model not only comes from new gauge bosons $W_{H}$
but also comes from correction to the couplings vertices of SM gauge
boson and Higgs boson. For the fixed $f$, c and $c'$, the correction
cross section $\delta\sigma$ mainly proportional to the factor
$3xf/v$ at the order of $v^{2}/f^{2}$, which come from
 the
coupling vertices of $W_{L}W_{L}H$ in the LH model. As long as
$c>0.1$, the value of $\delta\sigma/\sigma^{SM}$ is larger than
$10\%$ in most of the parameter space of the LH model. On the other
hand, we can see that the value of $\delta\sigma/\sigma^{SM}$
decreases as $f$ increasing, which is consistent with the
conclusions for the
corrections of the LH model to other observables. \\
 \indent As has been mentioned above, the value of the
relative correction is larger than $10\%$ in most of the parameter
space preferred by the electrowesk precision data, the cross section
of $e^{-}\gamma\rightarrow \nu_{e}W^{-}H$ can amounts to about
$10^{3}$ events with the integrated luminosity of $100fb^{-1}$. The
$1\sigma$ statistical error corresponds to about $1\%$ precision.
So, such correction might be detected via $e\gamma$ collisions in
future $LC$ experiment with $\sqrt{s}$=500 GeV and
$\pounds=100fb^{-1}$.
\begin{figure}[h]
\begin{center}
\scalebox{0.85}{\epsfig{file=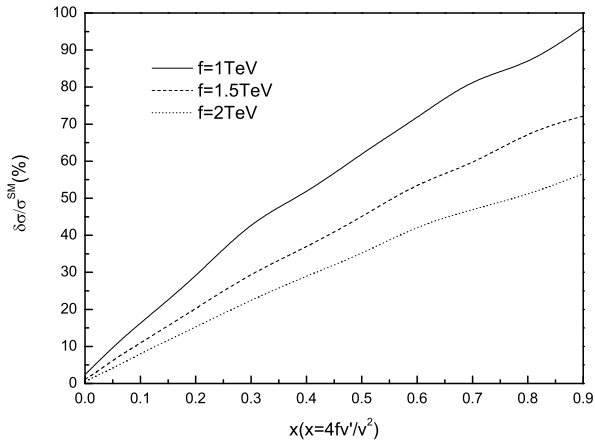}}
\end{center}
\caption{\small The relative correction $\delta\sigma/\sigma^{SM}$
as a function of the the parameter x($x=4fv'/v^{2}<1$) for $c=0.3$,
$c'=0.68$ and three values of the scalar parameter $f$.}
\end{figure}

\section{Conclusion}
 \indent The little Higgs model, which
can solve the hierarchy problem, is a promising alternative model of
new physics beyond the standard model. Among the various little
Higgs models, the littlest Higgs model is one of the simplest and
phenomenologically viable models. The distinguishing feature of this
model is the existence of the new scalars, the new gauge bosons, and
the vector-like top quark. These new particles contribute to the
experimental observables, which could provide some clues of the
existence of the littlest Higgs model. In this paper, we study the
potential to detect the contribution of the littlest Higgs model via
the process $e^{-}\gamma\rightarrow \nu_{e}W^{-}H$ at the future
$LC$ experiments.

\indent In the parameter spaces($f=1\sim2$ TeV, $c=0\sim0.5$,
$c'=0.62\sim0.73$) limited by the electroweak precision data, we
calculate the cross section correction of the littlest Higgs model
to the process $e^{-}\gamma\rightarrow \nu_{e}W^{-}H$. We find that
the correction is significant even when we consider the constraint
of electroweak precision data on the parameters. In most of
parameter space, the relative correction can be over $10\%$, which
can be seen as the new signals of light Higgs boson and should be
detected via this process at the future LC experiment. The littlest
Higgs model
 is a weak interaction theory and it
is hard to detect its contributions and measures its couplings at
the LHC. With the high c.m. energy and luminosity, the future $LC$
experiment will open an ideal window to probe into the littlest
Higgs model and study its properties. With the relative correction
over $10\%$ of the littlest Higgs model, we believe that the process
$e^{-}\gamma\rightarrow \nu_{e}W^{-}H$ can provide us significant
signal of the LH model at future $LC$ experiment.

\end{document}